\documentstyle[amsmath,bbm,amssymb,amsfonts,epsfig,12pt,here]{article}

\newcommand {\slsh} [1] {\not{\hbox{\kern-2pt${#1}$}}}

\def\drawbox#1#2{\hrule height#2pt
         \hbox{\vrule width#2pt height#1pt \kern#1pt
               \vrule width#2pt}
               \hrule height#2pt}

\def\Asym#1#2{\vcenter{\vbox{\drawbox{#1}{#2}
               \kern-#2pt       
               \drawbox{#1}{#2}}}}

\newcommand {\beq} {\begin{equation}}
\newcommand {\eeq} {\end{equation}}
  \newcommand {\ber}{\begin{eqnarray*}}
  \newcommand {\eer} {\end{eqnarray*}}
\newcommand {\bea}{\begin{eqnarray}}
  \newcommand {\eea} {\end{eqnarray}}

\newcommand{\None}{${\cal N}=1\ $}

\newcommand{\Dslash}{\,{\raise.15ex\hbox{/}\mkern-12mu D}}

\def\Acknowledgements{\bigskip  \bigskip {\begin{center} \begin{large}
              \bf ACKNOWLEDGMENTS \end{large}\end{center}}}

\begin{document}
\begin{titlepage}

\vskip 1cm

\centerline{{\Large \bf The Hagedorn Temperature}}
\vskip 0.4mm
\centerline{{\Large \bf and Open QCD-String Tachyons }}
\vskip 0.4mm
\centerline{{\Large \bf in Pure \None Super Yang-Mills}}

\vskip 1cm
\centerline{\large Adi Armoni and Timothy J. Hollowood}

\vskip 0.5cm
\centerline{Department of Physics, Swansea University}
\centerline{Singleton Park, Swansea, SA2 8PP, UK}

\vskip 1cm

\begin{abstract}

We consider large-$N$ confining gauge theories with a Hagedorn density of states. In such theories the potential between a pair of colour-singlet sources may diverge at a critical distance $r_c = 1/ T_H$. We consider, in particular,  pure \None super Yang-Mills theory and argue that when a domain-wall and an anti domain-wall are brought to a distance near $r_c$ the interaction potential is better described by an ``open QCD-string channel''. We interpret the divergence of the potential in terms of a tachyonic mode and relate its mass to the Hagedorn temperature. Finally we relate our result to a theorem of Kutasov and Seiberg and argue that
the presence of an open string tachyonic mode in the annulus amplitude implies an exponential density of states in the UV of the closed string channel.

\end{abstract}

\end{titlepage}

\section{Introduction}
\label{introduction}

\noindent
 Gauge field theories with an asymptotic exponential density of states
 admit a ``limiting temperature'' called the Hagedorn temperature (see \cite{Barbon:2004dd} for a recent review).
 It is believed that four dimensional gauge theories such as pure Yang-Mills (or pure \None super Yang-Mills) confine and that the colour-singlet states exhibit a Hagedorn behaviour. The goal of this note is to explore the implications
of the Hagedorn behaviour on the short distance, 
$r<\Lambda _{\rm QCD}^{-1}$, dynamics of confining theories.

Our main focus is large-$N$ pure \None super Yang-Mills, since in this
theory the vacuum structure of the theory is well understood: the
$U_R(1)$ symmetry is broken by the anomaly to a discrete ${\bf Z}_{2N}$
symmetry which is further broken down spontaneously to ${\bf Z}_2$
\cite{Witten:1982df,Veneziano:1982ah}. The gluino condensate is the
order parameter for this breaking \cite{Shifman:1987ia} which labels
the $N$ vacuum states. The fact that there are degenerate vacua means
that there are domain walls in this theory which separate regions of
different vacuum. What is particularly special is that planar domain
walls are BPS objects that preserve one half of the supersymmetry of
the theory \cite{Dvali:1996xe}. These walls will be of prime interest to us.

\None SYM is in some respects a good toy model for real QCD. The model
consists of one quark flavor in the adjoint representation and its
relation to ordinary one flavor QCD is not only qualitative. The model
is expected to confine and the glueball spectrum is expected to have a
mass gap. Understanding this model in full is an outstanding problem in
its own right which will likely teach us much about real QCD. However,
for present purposes we should note that QCD does not have domain walls.

Returning to the supersymmetric theory, we
will consider a setup of a domain wall and an anti-domain wall and
argue that as the walls are brought close to each other, the system
develops a tachyonic QCD-string mode. This is in keeping with the
interpretation of the domain wall as a D-brane \cite{Witten:1997ep}.
Our main observation is a relation between the Hagedorn temperature 
and the mass of the tachyon
\beq
|M_0| = {\sigma \over T_H}\, ,
\eeq
where $M_0$ is the tachyon mass, $T_H$ is the Hagedorn temperature and $\sigma$ is the QCD-string tension.

The paper is organized as follows: in Section 2 we discuss the general behaviour of a potential between colour-singlets in confining theories. In Section 3 we focus on domain walls in \None SYM. Section 4 is devoted to a generalization of a theorem by Kutasov and Seiberg. 

\section{The Potential Between Colour-Singlets}

Consider two colour singlets in a confining gauge theory, separated by
a distance $r$. The colour singlets may be either glueballs, or QCD-strings (Wilson-loops) or domain walls.

The sources interact via the exchange of glueballs. If the source
itself is a glueball or a QCD-string, the interaction is suppressed by
$1/N$. If the source is a domain wall the interaction is not
suppressed even at infinite $N$. At large enough $N$, the interaction
is dominated by a tree-level exchange of glueballs. Therefore a
general expression for the potential between the sources is 
\beq
V(r)\thicksim \sum _n |C_n| ^2 d_n \int d^3k \frac{e^{ikr}} {k^2 + M^2_n} \thicksim \sum _n |C_n|^2 d_n
e^{-M_n r} \, . \label{V1} 
\eeq
Here, the sum, is over glueballs states and $d_n$ is the number of 
glueballs of mass $M_n$. $C_n$ is the glueball-source coupling. Our discussion below will be valid in cases where $C_n$ does not depend exponentially on $n$. An exponential dependence is not expected to occur in the wall-source case, as suggested by the type II annulus amplitude. On the other hand it does occur in the Veneziano amplitude.\footnote{We thank A. Schwimmer for bringing this example to our attention.} Clearly, the interaction at large separation is controlled by the
lowest states in the tower. 
If the system exhibits an asymptotic Hagedorn behaviour 
$d_n \rightarrow \exp \gamma M_n$, for a constant $\gamma$, 
the potential \eqref{V1} then takes the form
\beq
V(r)\thicksim \sum _n e^{\gamma M_n}\,e^{-M_n r} \, . \label{V2} 
\eeq 
We will ignore any sub-leading polynomial in $M_n$ that can multiply
$e^{\gamma M_n}$. Notice that \eqref{V2} has the form of the canonical
  partition function of the system at a temperature $T=1/r$. In
  particular, the constant $\gamma$ is identified with with the
  inverse Hagedorn temperature, $\gamma=1/T_H$. Now the canonical
  partition function is not well defined when $T\geq T_H$, and at the
  Hagedorn temperature $T=T_H$ the partition function diverges. This
  implies that the potential \eqref{V2} diverges when $r$ is reduced
to a critical distance $r_c=1/T_H$.
What is the physical significance of this divergence? 
In general, it means that system is not stable and that there exists a
better description of the interaction in terms of different degrees of
freedom. When the sources are glueballs or QCD-strings, the answer is
that the potential should be calculated by using the short distance
degrees of freedom---the gluons, which interact with the constituents of
the source. The subject of the next section is to discuss the meaning of the
divergence in the potential when the sources are domain walls. 

Before we continue, it is useful to  
replace the sum 
over the glueballs states in \eqref{V2} by an integral over a mass density:
\beq
V(r)\thicksim \int dM\, e^{\gamma M}\,e^{-Mr} \,. \label{V3}
\eeq
   
\section{The Domain Wall Anti-Domain Wall System}
\noindent

\None SYM $SU(N)$ theory admits $N$ degenerate vacua. Hence there
exist domain walls that separate those vacua \cite{Dvali:1996xe}. The
tension of the walls is given by the absolute value of the difference
between the values of the gluino condensate \cite{Dvali:1996xe} 
\beq
T_k = {N^2 \Lambda _{\rm QCD} ^3 \over 4\pi ^2} \sin {\pi k \over N} \, .
\eeq  
The tension of a fundamental wall is $\sim N$. In string theory the
tension of a D-branes is $T\sim 1/g_{\rm st}$. Together with the
identification $g_{\rm st} \sim 1/N$, it is suggestive of the
conjecture that the 
domain walls are D-branes for the \None QCD string \cite{Witten:1997ep}. 

There is further evidence that domain walls are D-branes. First of
all, the QCD string can end on a domain wall
\cite{Witten:1997ep,Shifman:2002jm}. Domain walls interact via an
exchange of glueballs \cite{Armoni:2003jk} (states of the ``QCD closed
string''). Moreover, it has been argued that the collective dynamics
of a stack of domain walls is described by a 3d gauge theory 
on the world volume \cite{Acharya:2001dz}. By using this 3d world-volume gauge theory an explicit two-loops calculation of the interaction between two domain walls was made in \cite{Armoni:2005sp,Armoni:2006ee}. This is a field theory example of open/closed string channel duality.  

Let us briefly review how two domain walls interact. The domain walls carry a tension and a charge (in parallel to the NS-NS tension and R-R charge of D-branes in type II string theory). Even parity glueballs couple to the tension density $F^2$ and odd parity glueballs couple to the charge density $F\tilde F$. Supersymmetry implies the following identity \cite{Armoni:2003jk}
\beq
0= \int d^4 x \, \left ( \langle F^2 (x), F^2 (0) \rangle - \langle F\tilde F (x), F \tilde F (0) \rangle \right ) \, .\label{int}
\eeq 
This is saturated at large-$N$, by the exchange of even and odd parity glueballs, namely
\bea
& & 
 \langle F^2 (x), F^2 (0) \rangle - \langle F\tilde F (x), F \tilde F (0) \rangle = \nonumber \\
& & \int d^4 q\,e^{iq\cdot x} \, \left (\sum _ + {f_n^2 \over {q^2 + M_n ^2}} - \sum _ - {f_n^2 \over {q^2 + M_n ^2}}  \right ) \, . \label{glue}
\eea 
The vanishing of the right-hand side of this equation is due to supersymmetry: the couplings $f_n$ and the masses $M_n$ of the even and odd parity glueballs are degenerate. The interpretation is clear: if we place two parallel domain walls there will be no force between them, since this is a BPS configuration (at large-$N$). The microscopic reason for the vanishing of the force is a perfect cancellation between the two glueball towers.

Consider now the following set-up: a domain wall and an anti
domain-wall, separated by a certain distance $r$. To the left and to
the right of the configuration there exists the same vacuum
state. However, in between the walls the vacuum is different. The
walls are expected to attract each other and finally
annihilate. Intuitively, there is a probability of 
a tube of the first vacuum state being formed at
some local position on the domain walls. This tube would then
expand. Going back to the expression 
\eqref{glue}. The only difference is a sign: both the even parity glueballs and the odd parity glueballs attract
\bea
& & 
  \langle F^2 (x), F^2 (0) \rangle + \langle F\tilde F (x), F \tilde F (0) \rangle = \nonumber \\
& & \int d^4 q\, e^{iq\cdot x} \, \left (\sum _ + {f_n^2 \over {q^2 + M_n ^2}} + \sum _ - {f_n^2 \over {q^2 + M_n ^2}}  \right ) \, . \label{glue2}
\eea

The potential \eqref{V3} between a wall and an anti-wall can be written as 
\beq
V(r) \thicksim \int _0 ^\infty ds \int dk \int M^2 dM \, e^{M/T_H}\,
e^{-s(k^2 +M^2)}\,e^{ikr} \label{V31} \, .
\eeq
The integration over $k$ and $M$ yields 
 \beq
V(r) \thicksim \int ds\, e^{1/(4s T_H^2)}\,e^{-r^2/4s}
 \label{V4} \, .
\eeq 
Written in terms of the variable $t=1/(4\sigma ^2s)$ the potential takes the form
 \beq
V(r) \thicksim \int dt \,e^{ t(\sigma ^2/T_H^2 - (\sigma r)^2)}
 \label{V5} \, .
\eeq

Assuming that there is a field theory living on the wall anti-wall system, then we can calculate the potential by the ``open string channel'', namely by calculating the Casimir energy of the system.
The contribution of a single massive state to the Casimir energy is
\beq
E(r) = \int {d^3 k \over (2\pi) ^3} \log (k^2 + M^2)= \int {dt \over t} {1 \over t^{3/2}}\, e^{-t M^2} \, .\label{t}
\eeq
The energy \eqref{t} can be matched with the potential \eqref{V5} if
\beq
M^2 = - M_0^2 + (\sigma r)^2 \, , \label{tachyon}
\eeq
with 
\beq
M_0^2 = {\sigma ^2 \over T_H^2} \, . \label{r1}
\eeq
Namely, there must exists a tachyonic mode in the ``open QCD-string
channel'' whose mass respects \eqref{tachyon} and \eqref{r1}.

Our result is an interesting UV/IR relation within field theory: we related the Hagedorn temperature - a property of the the UV to a tachyonic mass (IR mode). Note that the above derivation assumes a stringy picture, but it does not assume a particular string theory.

\section{A Relation Between Open Strings and Closed Strings}
\noindent

In this section we wish to comment on the relation between our field 
theory result and similar relations in string theory. 

Kutasov and Seiberg proved a while ago \cite{Kutasov:1990sv}, by using modular invariance, the following relation in any oriented closed string theory:
if the density of states of bosons minus the density of states of fermions is exponential, it implies a tachyon. This is a remarkable UV/IR relation.
The theorem was extended to open string theory by Niarchos \cite{Niarchos:2000kw} who showed by analyzing the annulus diagram that a closed string tachyon is related to an exponential density of states in the open string channel.

We wish to further extend the theorem to a relation between an open string tachyon and a Hagedorn density of NS-NS (or R-R) states. A related discussion for critical type II string theory can be found in refs.\cite{Gaiotto:2003rm,Sarangi:2003sg}.

Consider the annulus amplitude in open string theory. It has the following
structure
\beq
\int {dt \over t} A(t) \, ,
\eeq
where $A(t)$ is the vacuum energy of the open string tower
\beq
A(t) = \int {d^D p \over (2\pi)^D} \sum _ne^{-t(p^2 + M_n ^2)}   \, .
\eeq
The contribution from the lowest state of the tower is simply
\beq
A(t) \sim {1 \over t^{D/2}} e^{-t M_0 ^2}
\eeq
and in particular if the lowest state is tachyonic $M_0^2 <0$, the annulus amplitude diverges at $t\rightarrow \infty$. This is an IR divergence.

There is another way of understanding this divergence. The same amplitude can be written in terms of the variable $\tau = {1\over t}$,
\beq
\int {d\tau \over \tau } A(\tau) \, ,
\eeq
with
 \beq
A(\tau)=  \tau ^{D/2} e^{|M_0 ^2 |/ \tau} \, \label{exp} .
\eeq
The interpretation of the amplitude, written in terms of the variable $\tau$, is of a propagation of closed string between a stack of two branes. The behaviour at $\tau \rightarrow 0$ is a property of the UV regime. The exponential divergence at the the UV \eqref{exp} should be interpreted as an exponential density of bosonic (NS-NS or R-R) closed string states, exactly as in \cite{Kutasov:1990sv}. 
Thus we found the following relation between the IR of the open string theory and the UV of the closed string theory: an open string tachyon exists if and only if the asymptotic density of closed strings that propagate in the annulus diagram is exponential. 

Let us consider one particular case. A system of a brane and an anti-brane in type II string theory.
When the branes are brought close to each other a tachyonic mode develops, since the mass of the lightest open string modes is $\alpha ' M^2 = -{1\over 2} + {y^2 \over 4 \pi ^2 \alpha' }$. Physically, this is due to the instability of the system: a lower energy state can be achieved by the annihilation of the branes. The explanation in terms of the closed string channel is also straightforward: the NS-NS and R-R sectors contribute equally and {\it with equal signs}. Thus one can multiply the NS-NS contribution by two. The amplitude exhibits an asymptotic exponential density of states. This is perhaps the simplest example of a bosonic string amplitude that exhibits our claimed IR/UV connection.

\Acknowledgements

A.A. wishes to thank E. Imeroni, N. Itzhaki, D. Kutasov, B. Lucini, A. Naqvi, V. Niarchos and A. Schwimmer for useful discussions. Special thanks to M. Shifman for comments on an early draft of the paper and for discussions. A.A. is supported by the PPARC advanced fellowship award.


\begin{thebibliography}{99}

\bibitem{Barbon:2004dd}
  J.~L.~F.~Barbon and E.~Rabinovici,
  ``Touring the Hagedorn ridge,''
  arXiv:hep-th/0407236.

\bibitem{Witten:1982df}
  E.~Witten,
  ``Constraints On Supersymmetry Breaking,''
  Nucl.\ Phys.\  B {\bf 202}, 253 (1982).

\bibitem{Veneziano:1982ah}
  G.~Veneziano and S.~Yankielowicz,
  ``An Effective Lagrangian For The Pure N=1 Supersymmetric Yang-Mills Theory,''
  Phys.\ Lett.\  B {\bf 113}, 231 (1982).

\bibitem{Shifman:1987ia}
  M.~A.~Shifman and A.~I.~Vainshtein,
  ``On Gluino Condensation in Supersymmetric Gauge Theories. SU(N) and O(N) Groups,''
  Nucl.\ Phys.\  B {\bf 296}, 445 (1988)
  [Sov.\ Phys.\ JETP {\bf 66}, 1100 (1987)].

\bibitem{Dvali:1996xe}
  G.~R.~Dvali and M.~A.~Shifman,
  ``Domain walls in strongly coupled theories,''
  Phys.\ Lett.\  B {\bf 396}, 64 (1997)
  [Erratum-ibid.\  B {\bf 407}, 452 (1997)]
  [hep-th/9612128].

\bibitem{Witten:1997ep}
  E.~Witten,
  ``Branes and the dynamics of {QCD},''
  Nucl.\ Phys.\  B {\bf 507}, 658 (1997)
  [hep-th/9706109].

\bibitem{Shifman:2002jm}
  M.~Shifman and A.~Yung,
  ``Domain walls and flux tubes in N = 2 SQCD: D-brane prototypes,''
  Phys.\ Rev.\  D {\bf 67}, 125007 (2003)
  [hep-th/0212293].

\bibitem{Acharya:2001dz}
  B.~S.~Acharya and C.~Vafa,
  ``On domain walls of N = 1 supersymmetric Yang-Mills in four dimensions,''
hep-th/0103011.

\bibitem{Armoni:2003jk}
  A.~Armoni and M.~Shifman,
  ``The cosmological constant and domain walls in orientifold field  theories
  and N = 1 gluodynamics,''
  Nucl.\ Phys.\  B {\bf 670}, 148 (2003)
  [hep-th/0303109].

\bibitem{Armoni:2005sp}
  A.~Armoni and T.~J.~Hollowood,
  ``Sitting on the domain walls of N = 1 super Yang-Mills,''
  JHEP {\bf 0507}, 043 (2005)
  [hep-th/0505213].

\bibitem{Armoni:2006ee}
  A.~Armoni and T.~J.~Hollowood,
  ``Interactions of domain walls of SUSY Yang-Mills as D-branes,''
  JHEP {\bf 0602}, 072 (2006)
  [hep-th/0601150].

\bibitem{Kutasov:1990sv}
  D.~Kutasov and N.~Seiberg,
  ``Number Of Degrees Of Freedom, Density Of States And Tachyons In String Theory And Cft,''
  Nucl.\ Phys.\  B {\bf 358}, 600 (1991).

\bibitem{Niarchos:2000kw}
  V.~Niarchos,
  ``Density of states and tachyons in open and closed string theory,''
  JHEP {\bf 0106}, 048 (2001)
  [hep-th/0010154].

\bibitem{Gaiotto:2003rm}
  D.~Gaiotto, N.~Itzhaki and L.~Rastelli,
  ``Closed strings as imaginary D-branes,''
  Nucl.\ Phys.\  B {\bf 688}, 70 (2004)
  [hep-th/0304192].

\bibitem{Sarangi:2003sg}
  S.~Sarangi and S.~H.~H.~Tye,
  ``Inter-brane potential and the decay of a non-BPS-D-brane to closed strings,''
  Phys.\ Lett.\  B {\bf 573}, 181 (2003)
  [arXiv:hep-th/0307078].
\end{thebibliography}
\end{document}